\documentclass[preprint,12pt,3p]{elsarticle}
\usepackage{graphicx}
\usepackage{dcolumn}
\usepackage{amssymb}
\usepackage{epsfig}
\usepackage{amsthm}
\usepackage{amsmath}
\usepackage{mathrsfs}
\usepackage{float}
\usepackage{breqn}

\begin{document}   
\journal{Chem. Phys.}
\title{Exact results on diffusion in a piecewise linear potential with a rectangular sink}
\author[label1]{Proma Mondal\corref{cor1}}
\cortext[cor1]{Corresponding author}
\ead{promamon1992@gmail.com}
\author[label2]{Aniruddha Chakraborty}
\address{School of Basic Sciences, Indian Institute of Technology Mandi, Kamand, Himachal Pradesh-175005, India}
\begin{frontmatter}
\begin{abstract}
We propose a new  method for finding the exact analytical solution in Laplace domain for the problem where the probability density of a random walker in a piece-wise linear potential in presence of a rectangular sink of arbitrary width and height. The motion of the  random walker is modelled by using  Smoluchowski equation. For our model we have derived exact analytical expression for rate constants. This is the first model where the exact analytical solution in closed form is possible in the case of a sink of arbitrary width for position dependent potential. This model is better for understanding reaction-diffusion systems than all other existing models available in literature.
\end{abstract}
\date{\today}
\end{frontmatter}
\section{Introduction}
\noindent Reaction-diffusion system is of significant importance to the research community. \cite{Risken,Nishijima,Agmon,Dobler,Mathies,Hong,Archer,Luczkaa,Ansari,Szabo,Dagdug}. The standard method of modeling reaction diffusion system is done by using Smoluchowski equation with an additional sink term. \cite{Nishijima,Agmon,Dobler,Mathies,Hong,Archer,Luczkaa,Ansari,Szabo,Dagdug}. Exact analytical results of this modified Smoluchowski equation helps in understanding the importance of different parameters like friction and provide an insight to different approximations. Sinks of different shapes has been used by different research groups for modelling different reaction-diffusion system {\it e.g.}, Pinhole, Gaussian \cite{Bagchi2}, and Lorentzian \cite{Bagchi} Dirac delta function sink \cite{Sebastian1,Sebastian2,Chakravarti} and ultrashort sink \cite{Himani}. For a Brownian particle it is possible to diffuse out from a potential well, if the  well is finite or if there is any sink \cite{SKG,Kenkre}.
One interesting examples of reaction-diffusion system is, reactions in condensed phase, {\it e.g.,} looping of a polymer in solution \cite{Moumita}, electron transfer reactions in DNA \cite{Rajarshi} and electronic relaxation of a molecule in solution \cite{Bagchi2,Bagchi,Samanta,Sebastian1}. A molecule inside a polar solvent is put on an electronically excited state potential energy surface by applying light of appropriate frequency. As lot of solvent molecules are present around that molecule, the molecular configuration executes a random walk on the electronically excited state potential energy surface. As the molecular configuration changes, it may undergo radiative decay from anywhere on that surface with equal rate. This radiative decay is theoretically modelled by adding an appropriate position independent term to the standard Smoluchowski equation.  The molecule may also undergo radiative decay from certain specified regions of that potential energy surface. Non-radiative decay is theoretically modelled by adding a position dependent sink term to the  Smoluchowski equation. In the following we propose a new sink of rectangular shape and find the exact analytical solution of the Smoluchowski equation in the case of a piece-wise linear potential. The Smoluchowski equation for the piece-wise linear potential is used earlier by Samanta {\it ety. al.,} \cite{SKG, Kenkre}. But in reality point sink is not the best model for most of the reaction-diffusion systems \cite{Bagchi1} but the the problem is exact analytical solution of Smoluchowski equation with a sink of finite width for any position dependent potential is not possible in closed form. The sink we propose in this paper is a rectangular sink and still we are able to solve the corresponding Smoluchowski equation analytically in Laplace domain for the case of piece-wise linear potential. In one dimensional case, the Smoluchowski equation for piece-wise linear potential with an arbitrary position dependent sink is given by
\begin{equation}
\frac{\partial P(x,t)}{\partial t} = D \left(\frac {\partial } {\partial x}\right)\left[\frac{\partial P(x,t)}{\partial x} + P(x, t) \left(\frac {\partial U}{\partial x}\right) \right] - k_r P(x,t) - S(x) P(x,t)
\end{equation}
This equation gives us the time dependent probability function $P(x,t)$ where the random walker is at the position $x$ at time $t$. $U(x)$ is the confining potential and $D$ is the diffusion coefficient. In this paper we have used piece-wise linear potential to represent the confining potential. In our previous paper, we have proposed a new model for reaction-diffusion system using a flat potential with a rectangular sink of  arbitrary width \cite{Proma1}.

\section{Formulation of the problem}       

\noindent In the following we will solve the Smoluchowski equation modified by the addition of a sink term as shown in Fig. 1.
\begin{figure}
\centering 
{\includegraphics[width=90mm]{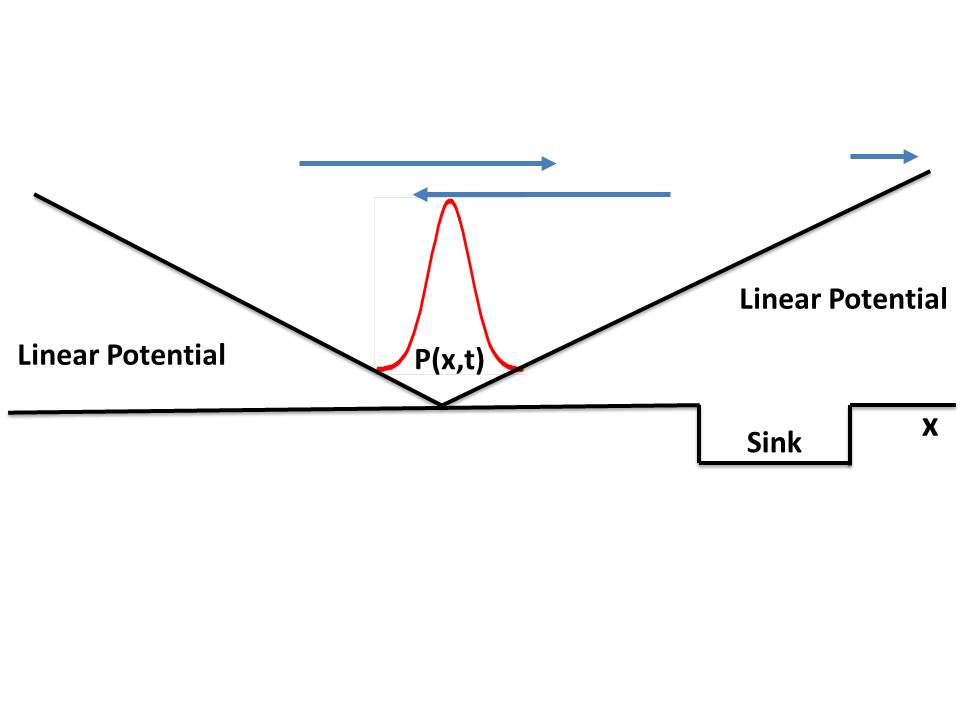}}\caption{
Schematic diagram showing the formulation of our problem.}
\end{figure}
In this paper we consider the case with piece-wise linear potential which is given by
\begin{eqnarray}
U(x) = -x\omega,\; -\infty \leq x \leq 0 \nonumber \\
U(x) = x\omega,\; 0 \leq x \leq \infty \nonumber \\
U(x) = \omega |x|.
\end{eqnarray}
\noindent Now Eq. (1) for piece-wise linear potential becomes
\begin{equation}
\frac{\partial P(x,t)}{\partial t} = D\frac{\partial}{\partial x}\left[\frac{\partial P(x,t)}{\partial x}+ \omega P(x,t) \;{sgn(x)} \right] - k_r P(x,t) -  S(x) P(x,t)
\end{equation}
\noindent Now we define Laplace transformation of $P(x,t)$, as given below
\begin{equation}
\tilde P(x,s)= \int^\infty_0 P(x,t) e^{-st} dt.
\end{equation}
Taking the Laplace transform of above equation gives
\begin{equation}
s \tilde{P}(x,s) - P(x,0) = D\left[\frac{\partial^2}{\partial x^2}\tilde{P}(x,s)+\text{sgn(x)} \omega\frac{\partial}{\partial x}\overline{P}(x,s)\right] - k_r \tilde P(x,s) - S(x) \tilde P(x,s),
\end{equation}
\noindent Now we assume $P(x,0)$ to be represented by a Dirac delta function. With this initial condition, Eq. (5) now becomes
\begin{equation}
s\tilde P(x,s)-\delta(x)= D\left[\frac{\partial^2}{\partial x^2}\tilde{P}(x,s)+\text{sgn(x)} \omega\frac{\partial}{\partial x}\overline{P}(x,s)\right] - k_r \tilde P(x,s) - S(x) \tilde P(x,s).
\end{equation}
In the following we will solve Eq. (6) in case of a rectangular sink of arbitrary width and height.

\section{Exact solution for rectangular sink}

\noindent Here we assume that the sink function $S(x)$ is a non-zero constant for a range of $x$-values from $x=a$ to $x=b$. Therefore we write $S(x) = V_{0} f(x)$, where $f(x)$ equals to $1$ for $x$ values between $a$ and $b$ and $f(x)$ equals to zero otherwise, where $a$ and $b$ both are finite positive number. The presence of $f(x)$ ensures that the sink function is a non-zero constant for an arbitrary  ranges of $x$ values between $a$ and $b$. Now we  replace the  $S(x) \tilde P (x,s))$ term of Eq. (6) by  $V_{0}\tilde P (x,s) f(x)$,  so that Eq. (6) is now modified as 
\begin{equation}
s\tilde P(x,s)-\delta(x)=  D\left[\frac{\partial^2}{\partial x^2}\tilde{P}(x,s)+\text{sgn(x)} \omega\frac{\partial}{\partial x}\overline{P}(x,s)\right] - V_0 \tilde P(x,s) f(x) - k_{r}\tilde P(x,s).
\end{equation} 
\noindent In the following Eq. (7) will be solved using the boundary condition method as outlined below. In region I, which is defined by $ - \infty \ge x > 0 $, Eq. (7) becomes
\begin{equation}
s\tilde P(x,s)=D \left[\frac{\partial^2\tilde P(x,s)}{\partial x^2}- \omega\frac{\partial}{\partial x}\overline{P}(x,s) \right] - k_{r}\tilde P(x,s).
\end{equation} 
\noindent The solution of the above equation is given by (with the assumption  $\tilde P(-\infty,s) = 0 $ )
\begin{equation}
\tilde P(x,s)= A e^{ (p+q) x}, 
\end{equation}
\noindent where 
\begin{eqnarray}
p = \sqrt{q^2 + \frac{s+k_r}{D}} \nonumber \\
q = \frac{\omega}{2}
\end{eqnarray}
In region II, which is defined by $ 0 < x > a $, Eq. (7) becomes
\begin{equation}
s\tilde P(x,s)=D \left[\frac{\partial^2\tilde P(x,s)}{\partial x^2}+\omega\frac{\partial}{\partial x}\overline{P}(x,s) \right] - k_{r}\tilde P(x,s).
\end{equation} 
Solution of the above equation is given by 
\begin{equation}
    \tilde P(x, s) = B e^{ (p-q) x} + C e^{ - (p+q)x}
\end{equation}
\noindent In region III, which is defined by $ a \ge x \le b $, Eq. (7) becomes
\begin{equation}
s\tilde P(x,s)=D \left[\frac{\partial^2\tilde P(x,s)}{\partial x^2}+\omega\frac{\partial}{\partial x}\overline{P}(x,s) \right] - V_0 \tilde P(x,s) - k_{r}\tilde P(x,s).
\end{equation} 
Solution of the above equation is given by
\begin{equation}
    \tilde P(x, s) = E e^{ (p_{1}-q) x} + F e^{ - (p_{1}+q)x},
\end{equation}
where
\begin{eqnarray}
p_{1} = \sqrt{q^2 + \frac{s+k_r+V_{0}}{D}} 
\end{eqnarray}
\noindent In region IV, which is defined by $ b < x \le \infty $, Eq. (7) becomes
\begin{equation}
s\tilde P(x,s)=D \left[\frac{\partial^2\tilde P(x,s)}{\partial x^2}+\omega\frac{\partial}{\partial x}\overline{P}(x,s) \right] - k_{r}\tilde P(x,s).
\end{equation} 
The solution of this equation (with the assumption  ($\tilde P(\infty,s) = 0 $ ) is given by 
\begin{equation}
\tilde P(x,s)= G e^{- (p+q) x}   
\end{equation}
From Eq. (7) it is obvious that probability has to be continuous at all values of x. This needs to following three boundary conditions which are given below:
\begin{eqnarray}
\tilde P(0 + \epsilon, s) & = \tilde P( 0 - \epsilon ,s), \nonumber \\
\tilde P(a + \epsilon, s) & = \tilde P( a - \epsilon ,s), \nonumber \\
\tilde P(b + \epsilon, s) & = \tilde P( b - \epsilon ,s).
\end{eqnarray}
\noindent The Eq. (5) may be integrated over $x$ for appropriate limits to get the following three boundary conditions.
\begin{eqnarray}
\left[\frac{\partial \tilde P(x , s)}{\partial x}\right]_{x=a+\epsilon} & = \left[\frac{\partial \tilde P(x, s)}{\partial x}\right]_{x=a-\epsilon} \nonumber \\ 
\left[\frac{\partial \tilde P(x , s)}{\partial x}\right]_{x=b+\epsilon} & = \left[ \frac{\partial \tilde P(x, s)}{\partial x}\right]_{x=b-\epsilon}
\nonumber \\
\left[\frac{\partial \tilde P(x , s)}{\partial x}\right]_{x=0+\epsilon}  - \left[\frac{\partial \tilde P(x, s)}{\partial x}\right]_{x=0-\epsilon} & + \omega \left[\tilde P(x,s)\right] _{x=0+\epsilon} - \omega \left[\tilde P(x,s)\right] _{x=0-\epsilon} = - 1/D.
\end{eqnarray}
Using these six boundary conditions, we have derived analytical expressions for all six unknown constants which are given below :
\begin{equation}
A= \frac{A_N}{A_D}    
\end{equation}
where 
\begin{dmath}
A_N= p^2 e {a (-p-q)+a ({p_1}-q)+b (-{p_1}-q)}-p^2 e^{a (-p-q)+a (-{p_1}-q)+b ({p_1}-q)}-{p_1}^2 e^{a (-p-q)+a ({p_1}-q)+b
   (-{p_1}-q)}+{p_1}^2 e^{a (-p-q)+a (-{p_1}-q)+b ({p_1}-q)}+p^2 e^{a (p-q)+a ({p_1}-q)+b (-{p_1}-q)}-p^2 e^{a (p-q)+a (-{p_1}-q)+b
   ({p_1}-q)}+{p_1}^2 e^{a (p-q)+a ({p_1}-q)+b (-{p_1}-q)}-{p_1}^2 e^{a (p-q)+a (-{p_1}-q)+b ({p_1}-q)}-2 p {p_1} e^{a (p-q)+a ({p_1}-q)+b
   (-{p_1}-q)}-2 p {p_1} e^{a (p-q)+a (-{p_1}-q)+b ({p_1}-q)}
\end{dmath}
and 
\begin{dmath}
A_D= 2 {D} \left(p^2 q e^{a (-p-q)+a ({p_1}-q)+b (-{p_1}-q)}-p^2 q e^{a (-p-q)+a (-{p_1}-q)+b ({p_1}-q)}-{p_1}^2 q e^{a (-p-q)+a ({p_1}-q)+b (-{p_1}-q)}+{p_1}^2 q e^{a (-p-q)+a (-{p_1}-q)+b ({p_1}-q)}+p^3 e^{a (p-q)+a ({p_1}-q)+b (-{p_1}-q)}-p^3 e^{a (p-q)+a (-{p_1}-q)+b
   ({p_1}-q)}-2 p^2 {p_1} e^{a (p-q)+a ({p_1}-q)+b (-{p_1}-q)}-2 p^2 {p_1} e^{a (p-q)+a (-{p_1}-q)+b ({p_1}-q)}+p^2 q e^{a (p-q)+a ({p_1}-q)+b
   (-{p_1}-q)}-p^2 q e^{a (p-q)+a (-{p_1}-q)+b ({p_1}-q)}+p {p_1}^2 e^{a (p-q)+a ({p_1}-q)+b (-{p_1}-q)}-p {p_1}^2 e^{a (p-q)+a (-{p_1}-q)+b
   ({p_1}-q)}+{p_1}^2 q e^{a (p-q)+a ({p_1}-q)+b (-{p_1}-q)}-{p_1}^2 q e^{a (p-q)+a (-{p_1}-q)+b ({p_1}-q)}-2 p {p_1} q e^{a (p-q)+a
   ({p_1}-q)+b (-{p_1}-q)}-2 p {p_1} q e^{a (p-q)+a (-{p_1}-q)+b ({p_1}-q)}\right)
\end{dmath}
\begin{equation}
B= \frac{B_N}{B_D},  
\end{equation}
with
\begin{dmath}
B_{N}= 2 {p_1} e^{a (-p-q)+a (-{p_1}-q)+a ({p_1}-q)} \left(-e^{b (-{p_1}-q)} \left((-p-q) e^{b (-p-q)}-\alpha  e^{b (-p-q)}\right)-e^{b (-p-q)} \left(\alpha  e^{b
   (-{p_1}-q)}-(-{p_1}-q) e^{b (-{p_1}-q)}\right)\right)-\left(-e^{a (-{p_1}-q)} \left(\alpha  e^{a (-p-q)}-(-p-q) e^{a (-p-q)}\right)-e^{a (-p-q)}
   \left((-{p_1}-q) e^{a (-{p_1}-q)}-\alpha  e^{a (-{p_1}-q)}\right)\right) \left(e^{a ({p_1}-q)} \left(-e^{b (-{p_1}-q)} \left((-p-q) e^{b (-p-q)}-\alpha 
   e^{b (-p-q)}\right)-e^{b (-p-q)} \left(\alpha  e^{b (-{p_1}-q)}-(-{p_1}-q) e^{b (-{p_1}-q)}\right)\right)-e^{a (-{p_1}-q)} \left(-e^{b ({p_1}-q)}
   \left((-p-q) e^{b (-p-q)}-\alpha  e^{b (-p-q)}\right)-e^{b (-p-q)} \left(\alpha  e^{b ({p_1}-q)}-({p_1}-q) e^{b ({p_1}-q)}\right)\right)\right)
\end{dmath}
and
\begin{dmath}
B_D= {D} \left(2 p^3 e^{a (p-q)+a (-{p_1}-q)+a ({p_1}-q)+b (-p-q)+b (-{p_1}-q)}-2 p^3 e^{a (p-q)+2 a (-{p_1}-q)+b (-p-q)+b ({p_1}-q)}-4 p^2 {p_1}
   e^{a (p-q)+a (-{p_1}-q)+a ({p_1}-q)+b (-p-q)+b (-{p_1}-q)}-4 p^2 {p_1} e^{a (p-q)+2 a (-{p_1}-q)+b (-p-q)+b ({p_1}-q)}+2 p^2 q e^{a (-p-q)+a
   (-{p_1}-q)+a ({p_1}-q)+b (-p-q)+b (-{p_1}-q)}+2 p^2 q e^{a (p-q)+a (-{p_1}-q)+a ({p_1}-q)+b (-p-q)+b (-{p_1}-q)}-2 p^2 q e^{a (-p-q)+2 a
   (-{p_1}-q)+b (-p-q)+b ({p_1}-q)}-2 p^2 q e^{a (p-q)+2 a (-{p_1}-q)+b (-p-q)+b ({p_1}-q)}+2 p {p_1}^2 e^{a (p-q)+a (-{p_1}-q)+a ({p_1}-q)+b
   (-p-q)+b (-{p_1}-q)}-2 p {p_1}^2 e^{a (p-q)+2 a (-{p_1}-q)+b (-p-q)+b ({p_1}-q)}-2 {p_1}^2 q e^{a (-p-q)+a (-{p_1}-q)+a ({p_1}-q)+b (-p-q)+b
   (-{p_1}-q)}+2 {p_1}^2 q e^{a (p-q)+a (-{p_1}-q)+a ({p_1}-q)+b (-p-q)+b (-{p_1}-q)}+2 {p_1}^2 q e^{a (-p-q)+2 a (-{p_1}-q)+b (-p-q)+b
   ({p_1}-q)}-2 {p_1}^2 q e^{a (p-q)+2 a (-{p_1}-q)+b (-p-q)+b ({p_1}-q)}-4 p {p_1} q e^{a (p-q)+a (-{p_1}-q)+a ({p_1}-q)+b (-p-q)+b
   (-{p_1}-q)}-4 p {p_1} q e^{a (p-q)+2 a (-{p_1}-q)+b (-p-q)+b ({p_1}-q)}\right)
\end{dmath}
\begin{equation}
C= \frac{C_N}{C_D}    
\end{equation}
where
\begin{dmath}
C_{N}= p^2 e^{a (p-q)+a ({p_1}-q)+b (-{p_1}-q)}-p^2 e^{a (p-q)+a (-{p_1}-q)+b ({p_1}-q)}+{p_1}^2 e^{a (p-q)+a ({p_1}-q)+b (-{p_1}-q)}-{p_1}^2 e^{a
   (p-q)+a (-{p_1}-q)+b ({p_1}-q)}-2 p {p_1} e^{a (p-q)+a ({p_1}-q)+b (-{p_1}-q)}-2 p {p_1} e^{a (p-q)+a (-{p_1}-q)+b ({p_1}-q)}
\end{dmath}
and
\begin{dmath}
C_D= 2 {D} \left(p^2 q e^{a (-p-q)+a ({p_1}-q)+b (-{p_1}-q)}-p^2 q e^{a (-p-q)+a (-{p_1}-q)+b ({p_1}-q)}-{p_1}^2 q e^{a (-p-q)+a ({p_1}-q)+b
   (-{p_1}-q)}+{p_1}^2 q e^{a (-p-q)+a (-{p_1}-q)+b ({p_1}-q)}+p^3 e^{a (p-q)+a ({p_1}-q)+b (-{p_1}-q)}-p^3 e^{a (p-q)+a (-{p_1}-q)+b
   ({p_1}-q)}-2 p^2 {p_1} e^{a (p-q)+a ({p_1}-q)+b (-{p_1}-q)}-2 p^2 {p_1} e^{a (p-q)+a (-{p_1}-q)+b ({p_1}-q)}+p^2 q e^{a (p-q)+a ({p_1}-q)+b
   (-{p_1}-q)}-p^2 q e^{a (p-q)+a (-{p_1}-q)+b ({p_1}-q)}+p {p_1}^2 e^{a (p-q)+a ({p_1}-q)+b (-{p_1}-q)}-p {p_1}^2 e^{a (p-q)+a (-{p_1}-q)+b
   ({p_1}-q)}+{p_1}^2 q e^{a (p-q)+a ({p_1}-q)+b (-{p_1}-q)}-{p_1}^2 q e^{a (p-q)+a (-{p_1}-q)+b ({p_1}-q)}-2 p {p_1} q e^{a (p-q)+a
   ({p_1}-q)+b (-{p_1}-q)}-2 p {p_1} q e^{a (p-q)+a (-{p_1}-q)+b ({p_1}-q)}\right)
\end{dmath}
\begin{equation}
E= \frac{E_{N}}{E_D}    
\end{equation}
where 
\begin{dmath}
E_{N}= p^2 e^{a (-p-q)+a (p-q)+b (-{p_1}-q)}-p {p_1} e^{a (-p-q)+a (p-q)+b (-{p_1}-q)}
\end{dmath}
and
\begin{dmath}
E_{D}=  {D} \left(p^2 q e^{a (-p-q)+a ({p_1}-q)+b (-{p_1}-q)}-p^2 q e^{a (-p-q)+a (-{p_1}-q)+b ({p_1}-q)}-{p_1}^2 q e^{a (-p-q)+a ({p_1}-q)+b
   (-{p_1}-q)}+{p_1}^2 q e^{a (-p-q)+a (-{p_1}-q)+b ({p_1}-q)}+p^3 e^{a (p-q)+a ({p_1}-q)+b (-{p_1}-q)}-p^3 e^{a (p-q)+a (-{p_1}-q)+b
   ({p_1}-q)}-2 p^2 {p_1} e^{a (p-q)+a ({p_1}-q)+b (-{p_1}-q)}-2 p^2 {p_1} e^{a (p-q)+a (-{p_1}-q)+b ({p_1}-q)}+p^2 q e^{a (p-q)+a ({p_1}-q)+b
   (-{p_1}-q)}-p^2 q e^{a (p-q)+a (-{p_1}-q)+b ({p_1}-q)}+p {p_1}^2 e^{a (p-q)+a ({p_1}-q)+b (-{p_1}-q)}-p {p_1}^2 e^{a (p-q)+a (-{p_1}-q)+b
   ({p_1}-q)}+{p_1}^2 q e^{a (p-q)+a ({p_1}-q)+b (-{p_1}-q)}-{p_1}^2 q e^{a (p-q)+a (-{p_1}-q)+b ({p_1}-q)}-2 p {p_1} q e^{a (p-q)+a
   ({p_1}-q)+b (-{p_1}-q)}-2 p {p_1} q e^{a (p-q)+a (-{p_1}-q)+b ({p_1}-q)}\right)
\end{dmath}
\begin{equation}
F= \frac{F_N}{F_D} 
\end{equation}
where 
\begin{dmath}
F_{N} = p^2 e^{a (-p-q)+a (p-q)+b (-{p_1}-q)}-p {p_1} e^{a (-p-q)+a (p-q)+b (-{p_1}-q)}
\end{dmath}
and
\begin{dmath}
F_D = {D} \left(p^2 q e^{a (-p-q)+a ({p_1}-q)+b (-{p_1}-q)}-p^2 q e^{a (-p-q)+a (-{p_1}-q)+b ({p_1}-q)}-{p_1}^2 q e^{a (-p-q)+a ({p_1}-q)+b
   (-{p_1}-q)}+{p_1}^2 q e^{a (-p-q)+a (-{p_1}-q)+b ({p_1}-q)}+p^3 e^{a (p-q)+a ({p_1}-q)+b (-{p_1}-q)}-p^3 e^{a (p-q)+a (-{p_1}-q)+b
   ({p_1}-q)}-2 p^2 {p_1} e^{a (p-q)+a ({p_1}-q)+b (-{p_1}-q)}-2 p^2 {p_1} e^{a (p-q)+a (-{p_1}-q)+b ({p_1}-q)}+p^2 q e^{a (p-q)+a ({p_1}-q)+b
   (-{p_1}-q)}-p^2 q e^{a (p-q)+a (-{p_1}-q)+b ({p_1}-q)}+p {p_1}^2 e^{a (p-q)+a ({p_1}-q)+b (-{p_1}-q)}-p {p_1}^2 e^{a (p-q)+a (-{p_1}-q)+b
   ({p_1}-q)}+{p_1}^2 q e^{a (p-q)+a ({p_1}-q)+b (-{p_1}-q)}-{p_1}^2 q e^{a (p-q)+a (-{p_1}-q)+b ({p_1}-q)}-2 p {p_1} q e^{a (p-q)+a
   ({p_1}-q)+b (-{p_1}-q)}-2 p {p_1} q e^{a (p-q)+a (-{p_1}-q)+b ({p_1}-q)}\right)
\end{dmath}
\begin{equation}
G= \frac{G_N}{G_D}
\end{equation}
where
\begin{dmath}
G_N = -2 p {p_1} e^{a (-p-q)+a (p-q)-b (-p-q)+b (-{p_1}-q)+b ({p_1}-q)}
\end{dmath}
and
\begin{dmath}
G_D = {D} \left(p^2 q e^{a (-p-q)+a ({p_1}-q)+b (-{p_1}-q)}-p^2 q e^{a (-p-q)+a (-{p_1}-q)+b ({p_1}-q)}-{p_1}^2 q e^{a (-p-q)+a ({p_1}-q)+b
   (-{p_1}-q)}+{p_1}^2 q e^{a (-p-q)+a (-{p_1}-q)+b ({p_1}-q)}+p^3 e^{a (p-q)+a ({p_1}-q)+b (-{p_1}-q)}-p^3 e^{a (p-q)+a (-{p_1}-q)+b
   ({p_1}-q)}-2 p^2 {p_1} e^{a (p-q)+a ({p_1}-q)+b (-{p_1}-q)}-2 p^2 {p_1} e^{a (p-q)+a (-{p_1}-q)+b ({p_1}-q)}+p^2 q e^{a (p-q)+a ({p_1}-q)+b
   (-{p_1}-q)}-p^2 q e^{a (p-q)+a (-{p_1}-q)+b ({p_1}-q)}+p {p_1}^2 e^{a (p-q)+a ({p_1}-q)+b (-{p_1}-q)}-p {p_1}^2 e^{a (p-q)+a (-{p_1}-q)+b
   ({p_1}-q)}+{p_1}^2 q e^{a (p-q)+a ({p_1}-q)+b (-{p_1}-q)}-{p_1}^2 q e^{a (p-q)+a (-{p_1}-q)+b ({p_1}-q)}-2 p {p_1} q e^{a (p-q)+a
   ({p_1}-q)+b (-{p_1}-q)}-2 p {p_1} q e^{a (p-q)+a (-{p_1}-q)+b ({p_1}-q)}\right)
\end{dmath}
These six unknown constants defines the complete solution of the problem {\it i.e.,} the analytical expression of $\tilde P(x,s)$. Therefore,  we have an explicit formula for $\tilde P(x,s)$, now we calculate the survival probability $P(t) = \int_{-\infty}^{\infty} P(x,t)dx$. It is possible to calculate the Laplace transform of $\tilde P(s)$ of $P(t)$ easily. Again $\tilde P(s)$ is related to $\tilde P(x,s)$ by $\tilde P(s) = \int_{-\infty}^{\infty} \tilde P(x,s) dx$. Hence, analytical formula for $\tilde P(s)$ is given by 
\begin{equation}
\tilde P(s)= - \frac{P_N}{P_D}
\end{equation}
where
\begin{dmath}
P_{N}= \frac{p (p+{p_1})^2 (p-{p_1}) e^{a p-a q+2 b {p_1}}}{\left(p^2-q^2\right) ({p_1}+q)}-\frac{2 p {p_1} (p-q) e^{a p+a {p_1}+b {p_1}-b
   q}}{{p_1}^2-q^2}+\frac{(p+{p_1})^2 e^{2 a p+2 b {p_1}}}{p+q}+\frac{2 p {p_1} e^{a (p+{p_1})+b ({p_1}-q)}}{p+q}+\frac{q e^{2 a {p_1}}
   \left({p_1}^2-p^2\right)}{q^2-p^2}-\frac{(p-{p_1})^2 e^{2 a (p+{p_1})}}{p+q}+\frac{p (p+{p_1}) (p-{p_1})^2 e^{a (p+2 {p_1}-q)}}{(p-q) (p+q)
   ({p_1}-q)}+\frac{q e^{2 b {p_1}} \left({p_1}^2-p^2\right)}{p^2-q^2}
\end{dmath}
and
\begin{dmath}
P_D = {D} \left(e^{2 b {p_1}} \left(q \left({p_1}^2-p^2\right)-e^{2 a p} (p+{p_1})^2 (p+q)\right)+(p-{p_1})^2 (p+q) e^{2 a (p+{p_1})}+q e^{2 a {p_1}}
   (p+{p_1}) (p-{p_1})\right)
\end{dmath}
\noindent The average rate constant may be found from $ {k_{I}}^{-1}= \tilde P(0)$. From Eq. (32) we get 
\begin{equation}
{k_{I}}^{-1}= \frac{P_N^{0}}{P_D^{0}}.    
\end{equation}
where
\begin{dmath}
P_{N}^{0}= 2 {V0} \left({k_r} \left({D} \sqrt{\frac{4 {k_r}}{{D}}+\omega ^2} \sqrt{\frac{4 ({k_r}+{V0})}{{D}}+\omega ^2}+{D} \omega ^2+4
   {V0}\right)+{D} {V0} \omega  \left(\sqrt{\frac{4 {k_r}}{{D}}+\omega ^2}+\omega \right)+4 {k_r}^2\right) e^{\frac{1}{2} a
   \left(\sqrt{\frac{4 {k_r}}{{D}}+\omega ^2}-\omega \right)+b \sqrt{\frac{4 ({k_r}+{V0})}{{D}}+\omega ^2}}-4 {D} {k_r} {V0}
   \sqrt{\frac{4 {k_r}}{{D}}+\omega ^2} \sqrt{\frac{4 ({k_r}+{V0})}{{D}}+\omega ^2}  e^{\frac{1}{2} \left(a \left(\sqrt{\frac{4
   ({k_r}+{V0})}{{D}}+\omega ^2}+\sqrt{\frac{4 {k_r}}{{D}}+\omega ^2}\right)+b \left(\sqrt{\frac{4 ({k_r}+{V0})}{{D}}+\omega ^2}-\omega
   \right)\right)}-4 {k_r} ({k_r}+{V0}) \left({D} \sqrt{\frac{4 {k_r}}{{D}}+\omega ^2} \sqrt{\frac{4 ({k_r}+{V0})}{{D}}+\omega
   ^2}+{D} \omega ^2+4 {k_r}+2 {V0}\right) e^{a \sqrt{\frac{4 {k_r}}{{D}}+\omega ^2}+b \sqrt{\frac{4 ({k_r}+{V0})}{{D}}+\omega ^2}}-2
   {V0} \left({k_r} \left(-{D} \sqrt{\frac{4 {k_r}}{{D}}+\omega ^2} \sqrt{\frac{4 ({k_r}+{V0})}{{D}}+\omega ^2}+{D} \omega ^2+4
   {V0}\right)+{D} {V0} \omega  \left(\sqrt{\frac{4 {k_r}}{{D}}+\omega ^2}+\omega \right)+4 {k_r}^2\right) e^{\frac{1}{2} a \left(2
   \sqrt{\frac{4 ({k_r}+{V0})}{{D}}+\omega ^2}+\sqrt{\frac{4 {k_r}}{{D}}+\omega ^2}-\omega \right)}+2 {D} {V0} \omega 
   ({k_r}+{V0}) \left(\sqrt{\frac{4 {k_r}}{{D}}+\omega ^2}+\omega \right) e^{a \sqrt{\frac{4 ({k_r}+{V0})}{{D}}+\omega ^2}}+4 {k_r}
   ({k_r}+{V0}) \left(-{D} \sqrt{\frac{4 {k_r}}{{D}}+\omega ^2} \sqrt{\frac{4 ({k_r}+{V0})}{{D}}+\omega ^2}+{D} \omega ^2+4
   {k_r}+2 {V0}\right) e^{a \left(\sqrt{\frac{4 ({k_r}+{V0})}{{D}}+\omega ^2}+\sqrt{\frac{4 {k_r}}{{D}}+\omega ^2}\right)}-2 {D}
   {V0} \omega  ({k_r}+{V0}) \left(\sqrt{\frac{4 {k_r}}{{D}}+\omega ^2}+\omega \right) e^{b \sqrt{\frac{4 ({k_r}+{V0})}{{D}}+\omega ^2}}
\end{dmath}
and
\begin{dmath}
P_D^{0} = 4 {D}^2 {k_r} ({k_r}+{V0}) \left(\sqrt{\frac{4 {k_r}}{{D}}+\omega ^2}+\omega \right) \left(\left(\frac{{V0} \omega }{2 {D}}-\frac{1}{8}
   \left(\sqrt{\frac{4 {k_r}}{{D}}+\omega ^2}+\omega \right) e^{a \sqrt{\frac{4 {k_r}}{{D}}+\omega ^2}} \left(\sqrt{\frac{4
   ({k_r}+{V0})}{{D}}+\omega ^2}+\sqrt{\frac{4 {k_r}}{{D}}+\omega ^2}\right)^2\right) e^{b \sqrt{\frac{4 ({k_r}+{V0})}{{D}}+\omega
   ^2}}+\frac{1}{8} \left(\sqrt{\frac{4 {k_r}}{{D}}+\omega ^2}+\omega \right) \left(\sqrt{\frac{4 {k_r}}{{D}}+\omega ^2}-\sqrt{\frac{4
   ({k_r}+{V0})}{{D}}+\omega ^2}\right)^2 e^{a \left(\sqrt{\frac{4 ({k_r}+{V0})}{{D}}+\omega ^2}+\sqrt{\frac{4 {k_r}}{{D}}+\omega
   ^2}\right)}-\frac{{V0} \omega  e^{a \sqrt{\frac{4 ({k_r}+{V0})}{{D}}+\omega ^2}}}{2 {D}}\right)
\end{dmath}
In the following section we will discuss behaviour of $k_{I}$ in details. 

\section{Results and Discussion}
\noindent From Eq. (42) we understand that the average rate constant depends on the parameter ``a", which is the distance between the peak of the initial probability distribution and the nearest edge of the sink. This average rate constant is also sensitive to both the depth ($V_0$) and the width of the rectangular sink {\it i.e.,} $(b-a)$. As usual the average rate constant depends on the diffusion constant ($D$) as well as, on the radiative decay rate constant.

\section{Conclusion}
\noindent In this paper we proposed rectangular sink model for reaction-diffusion system and with this new sink we are able to provide an exact analytical solution of the corresponding Smoluchowski equation for piece-wise linear potential in Laplace domain. This model is more general over all the existing point sink or narrow sink models available in literature. Our method of solution is very general and can be used to solve related problems.

\section{Acknowledgements}
\noindent One of the authors (A.C.) wants to thank IIT Mandi for PDA  and another author (P.M.) thanks IIT Mandi for HTRA scholarship.

\end{document}